\documentclass[lettersize,journal]{IEEEtran}
\usepackage{amsmath,amsfonts}
\usepackage{algorithmic}
\usepackage{algorithm}
\usepackage{tabularx}
\usepackage{array}
\usepackage[caption=false,font=normalsize,labelfont=sf,textfont=sf]{subfig}
\usepackage{textcomp}
\usepackage{stfloats}
\usepackage{url}
\usepackage{verbatim}
\usepackage{graphicx}
\usepackage{cite}

\usepackage{tikz}
\usetikzlibrary{arrows.meta,positioning,fit,calc}

\begin{document}
%

\title{RT-SHCUA: Real-Time Self-Hosted Computer-Use Agent for UAV Control}

\author{Di~Lu,~\IEEEmembership{Member,~IEEE,}
Bo Zhang,
Xiyuan Li,
Yongzhi~Liao,
Xuewen~Dong,~\IEEEmembership{Member,~IEEE,}
Yulong~Shen,~\IEEEmembership{Member,~IEEE,}
Zhiquan~Liu, ~\IEEEmembership{Senior Member,~IEEE,}
and~Jianfeng~Ma,~\IEEEmembership{Member,~IEEE}
\thanks{\textbullet\ Di Lu, Bo Zhang, Xiyuan Li, Yongzhi Liao, Xuewen Dong, and Yulong Shen are with the School of Computer Science and Technology, Xidian University, Xi’an, Shaanxi 710071, China, and also with the Shaanxi Key Laboratory of Network and System Security, Xi'an, Shaanxi, 710071, China. 
E-mail: \{dlu, xwdong\}@xidian.edu.cn; z17691248297@gmail.com; lixiyuan9196@outlook.com; liaoyongzhi1010@stu.xidian.edu.cn; ylshen@mail.xidian.edu.cn.\\
\textbullet\ Zhiquan Liu is with the College of Cyber Security, Jinan University, Guangzhou 510632, China. (E-mail: zqliu@vip.qq.com).\\
\textbullet\ Jianfeng Ma is with the School of Cyber Engineering, Shaanxi Key Lab of Network and System Security, Xidian University, Xi’an, Shaanxi, 710071, China. E-mail: jfma@mail.xidian.edu.cn.}%
\thanks{Manuscript received April 19, 2021; revised August 16, 2021.}}

\markboth{IEEE Journal Template,~Vol.~XX, No.~XX, Month~2026}%
{Author \MakeLowercase{\textit{et al.}}: RT-SHCUA: Real-Time Self-Hosted Computer-Use Agent for UAV Control}


\maketitle

\begin{abstract}
Natural-language control offers a promising interface for unmanned aerial vehicles (UAVs), but directly applying self-hosted computer-use agents (SHCUAs) to UAV control introduces a structural mismatch. SHCUAs are designed for interactive host-side tool use, where delayed agent iterations are often acceptable. UAV control, however, is coupled with continuously changing physical states, strict timing constraints, safety risks, and security accountability. A stale, unauthorized, or tampered agent decision may therefore lead to unsafe or untraceable vehicle behavior.

This paper proposes a real-time and security-oriented restructuring of SHCUA-based UAV control. Instead of allowing an SHCUA to directly issue flight commands, we transform its outputs into contract-bound UAV skill invocations with explicit timing, state, authority, fallback, and evidence semantics. Based on this abstraction, we design an architecture that separates semantic reasoning from onboard execution and security/safety enforcement. Slow cloud or edge reasoning is used for mission understanding, while onboard components validate and dispatch only timely, authorized, and state-consistent skills. Security-critical enforcement points can be protected by TEE-style or microcontroller isolation mechanisms without moving the full language agent or high-frequency flight-control loop into trusted components. Prototype evaluation shows that RT-SHCUA maintains bounded task-level responsiveness while supporting degraded handling, trusted admission, and auditable evidence preservation for SHCUA-mediated UAV actions.
\end{abstract}


\section{Introduction}

Natural-language control is becoming an important interface for unmanned aerial vehicles (UAVs) and other autonomous vehicles. Instead of using professional control interfaces, users can express mission-level intentions in natural language, for example inspection, tracking, evidence collection, or return conditions. Recent systems have shown that large language models and agentic frameworks can translate natural-language instructions into UAV commands or mission plans.

This development is part of a broader shift in AI-enabled cyber-physical systems (CPS) and the Internet of Things (IoT). Intelligent agents are increasingly expected to connect user intent with software, sensing infrastructure, and physical devices. Self-hosted computer-use agents (SHCUAs), such as OpenClaw-like agents, represent an important step in this shift because they provide a general execution model for iterative tool use. Extending this model from host-side computer use to UAVs and other autonomous vehicles is therefore part of the broader evolution toward agent-mediated control of AI-driven CPS and IoT environments.

However, this direction also changes the nature of the control problem.
Conventional SHCUAs are designed for host-side resources such as files, browsers, shell commands, and plugins.
These operations are usually discrete, weakly time-bounded, and recoverable.
When the same agentic execution model is connected to a UAV, the agent no longer affects only digital resources; it becomes an entry point to a physical system whose state evolves continuously.
This exposes a structural mismatch between the interactive, serial, and relatively slow SHCUA loop and the real-time requirements of UAV execution.
A command that is reasonable when generated may become stale by the time it reaches the vehicle.

Security is another unavoidable concern.
UAVs already have mature capabilities for mission execution, navigation, avoidance, and failsafe recovery.
Our goal is not to replace these capabilities.
Instead, the key question is how an open-ended SHCUA decision can be safely and timely connected to them.
Once an SHCUA becomes a UAV control interface, each requested action must be checked for authority, integrity, freshness, and auditability.
Thus, the problem is not only whether a UAV can execute a natural-language task, but also whether an SHCUA-generated decision is eligible to become a UAV-executable action.

This paper addresses the real-time and security challenges of SHCUA-mediated UAV control.
The key idea is not to accelerate the SHCUA loop until it satisfies UAV-level timing constraints.
Instead, we restructure how an SHCUA is allowed to influence UAV behavior.
The SHCUA remains responsible for natural-language understanding and task-level reasoning, but its outputs are not directly executed as flight commands.
They are compiled into contract-bound UAV skill invocations, each carrying timing, state, authority, fallback, and evidence semantics.
An onboard runtime dispatches only timely and state-consistent invocations, while a security and safety enforcement layer mediates authority, freshness, fallback, and evidence checks.

This design decouples slow semantic reasoning from fast vehicle execution.
The cloud or edge side can perform complex task understanding and mission planning, while onboard components retain the ability to execute time-sensitive skills, reject expired decisions, and trigger bounded fallback actions without waiting for the SHCUA.
At the same time, UAV control is not left to ordinary software guardrails alone.
Security-critical control points, such as skill authorization, freshness checking, and evidence generation, can be protected by trusted or isolated enforcement mechanisms on onboard, edge, or resource-constrained platforms.
In this way, the SHCUA is transformed from a direct controller into a constrained task-level proposer whose influence on the UAV is mediated, time-bounded, auditable, and recoverable.

The main contributions of this paper are as follows.
\begin{itemize}
    \item We identify the structural mismatch between SHCUA-style tool use and UAV control, showing that the conventional ``natural-language understanding--tool invocation--observation feedback'' loop is not suitable for direct use in real-time vehicle execution.

    \item We propose a contract-bound UAV skill invocation model that transforms free-form SHCUA tool calls into UAV operations with explicit timing, state, authority, fallback, and evidence semantics.

    \item We design an SHCUA-UAV architecture that separates task-level semantic reasoning from onboard execution and security/safety enforcement, so that slow language reasoning does not block time-critical vehicle behavior.

    \item We introduce a security-oriented enforcement path that supports authority control, command integrity, anti-replay protection, accountability, and trusted or isolated enforcement.

    \item We implement a prototype and evaluate task-level responsiveness, mission scaling, degraded execution, trusted-path consistency and overhead, and evidence append/audit cost.
\end{itemize}

\section{Background and Motivation}

This section introduces the two technical contexts that motivate this work: self-hosted computer-use agents and natural-language UAV control.
It then explains why simply connecting these two lines of work is insufficient.
The purpose of this section is not to present the solution, but to establish the structural reason why SHCUA-to-UAV control requires a redesigned execution model.

\subsection{Self-Hosted Computer-Use Agents}

Self-hosted computer-use agents extend language-model assistants from text generation to direct host-side interaction.
A typical SHCUA receives a natural-language instruction, reasons about the task, invokes host-side tools, observes execution results, and iteratively updates its plan.
The tool surface may include files, shell access, browser/application control, and plugins.

This execution model is effective for ordinary computer-use tasks because most host-side operations are discrete and weakly time-bounded.
For example, editing a file, opening a browser page, running a shell command, or calling a plugin can usually tolerate seconds of reasoning delay.
Failures may cause task interruption or data modification, but they usually do not immediately interact with physical dynamics.

The key limitation is that SHCUA tool use was not originally designed for real-time physical control.
The agent loop assumes that the agent can reason, invoke a tool, wait for an observation, and then decide the next step.
This assumption becomes problematic when the tool is no longer a digital resource but a physical vehicle whose state evolves continuously.

\subsection{Natural-Language UAV Control}

Natural-language UAV control aims to let users describe flight missions at a higher semantic level.
Instead of manually specifying waypoints or low-level commands, the user may ask the UAV to inspect an area, follow a target, photograph a structure, return home, or modify the mission according to environmental observations.

Existing systems typically connect language models to UAV command interfaces, mission planners, simulation environments, or common UAV middleware.
This demonstrates the feasibility of translating language into UAV-related actions.
However, much of this work focuses on whether the model can generate valid commands or task plans, rather than whether an agentic execution loop can safely and reliably control a real-time physical system.

For UAVs, natural-language control should be understood as task-level interaction, not direct participation in the flight-control loop.
The language model is useful for interpreting intent and generating mission structure, but time-critical control and emergency response must remain on the onboard side, close to the vehicle state and the flight controller.

\subsection{Structural Mismatch between SHCUAs and UAV Control}

There is a structural mismatch between SHCUA-style computer use and UAV control.

First, there is a latency mismatch.
A conventional SHCUA may take seconds to understand an instruction, generate a plan, invoke a tool, observe the result, and replan.
In UAV control, some decisions have very short validity windows.
An obstacle-avoidance action, a hover decision, or an emergency return command may become invalid if it is delayed.

Second, there is an operation-semantics mismatch.
A host-side tool call such as \texttt{edit\_file}, \texttt{run\_command}, or \texttt{open\_browser} mainly affects digital resources.
A UAV skill such as \texttt{fly\_to} or \texttt{emergency\_land} affects a physical vehicle.
Such operations must be interpreted together with vehicle state, environment state, timing constraints, and mission authority.

Third, there is a feedback mismatch.
SHCUAs usually receive discrete tool observations.
UAVs require continuous monitoring of position, velocity, battery level, communication quality, GPS status, obstacle distance, mission phase, and environmental uncertainty.
A single observation returned to the language agent is not sufficient to determine whether a physical action remains safe or valid.

Fourth, there is a security and accountability mismatch.
Once an SHCUA becomes a UAV control entry point, the problem is not only whether the generated trajectory is physically safe.
The system must also determine whether the agent is authorized to issue the command, whether parameters are bound to the approved intent, whether expired or replayed decisions are rejected, and whether the final vehicle behavior can be audited and traced.

These mismatches motivate the central design principle of this paper:
an SHCUA should not directly control a UAV through free-form tool invocation.
Instead, its outputs should be transformed into contract-bound vehicle skill invocations that can be checked, scheduled, authorized, audited, and degraded by the UAV-side runtime.

\section{Problem Statement and Design Goals}

This section defines the scope, assumptions, threat and failure model, and design goals.
The purpose is to make clear what problem this work solves and what it deliberately does not attempt to solve.

\subsection{Scope and Assumptions}

This work focuses on task-level UAV control through an SHCUA-style agent.
The user expresses mission intent in natural language.
The SHCUA performs semantic understanding and task-level reasoning.
The system then compiles the agent output into UAV skill invocations, which are validated and dispatched by the onboard runtime.

We assume the UAV platform exposes a skill library above the low-level flight controller.
Representative skills include takeoff, area inspection, return-to-home, and emergency landing.
The skills are not raw flight-controller commands.
They are bounded operations with preconditions, state constraints, timing requirements, and fallback semantics.

We do not attempt to replace the high-frequency flight-control loop.
Attitude control, motor control, low-level stabilization, and time-critical physical control remain the responsibility of the flight controller and existing real-time control stack.
Similarly, this work does not place large language models inside the hard real-time flight-control path.

The system contains the following logical components:
\begin{itemize}
    \item a user issuing natural-language mission instructions;
    \item an SHCUA runtime for semantic reasoning and task-level planning;
    \item a cloud or edge model for complex language understanding when available;
    \item a mission compiler that transforms agent plans into contract-bound skill invocations;
    \item an onboard runtime that performs state-aware dispatch and local execution;
    \item a security enforcement component that validates authority, freshness, and evidence;
    \item a safety supervisor that checks physical constraints and emergency conditions;
    \item a UAV skill library and flight controller for actual vehicle execution.
\end{itemize}

\subsection{Threat and Failure Model}

We distinguish safety and reliability failures from security threats.

Safety and reliability failures include environmental hazards, degraded vehicle state, sensor or communication anomalies, and model-side delays.
These failures may make a previously planned action unsafe or infeasible.
The system must therefore avoid waiting for slow semantic reasoning when an immediate onboard response is required.

Security threats arise because the SHCUA becomes a control interface to a physical platform.
The adversary may manipulate the agent context, induce unauthorized or tampered skill calls, replay expired decisions, or interfere with observations and logs.
A compromised or manipulated host-side runtime may also attempt to bypass authorization checks, alter commands after approval, suppress evidence, or misbind a user intent to a different UAV action.

The system does not assume that the SHCUA itself is fully trustworthy.
Instead, the SHCUA is treated as a task-level proposer whose outputs must be mediated by runtime enforcement.
Security-critical decisions, such as authorization, freshness validation, and evidence generation, should not rely solely on an ordinary unprotected process.

\subsection{Design Goals}

The design is guided by five goals.

\textbf{G1: Real-time responsiveness.}
Slow SHCUA reasoning must not block time-critical UAV behavior.
The UAV should retain onboard mechanisms for immediate state-dependent hold, return, or emergency responses.

\textbf{G2: Contract-bound skill execution.}
The SHCUA should not issue raw flight commands.
Its outputs should be transformed into UAV skill invocations that explicitly carry timing, state, authority, fallback, and evidence constraints.

\textbf{G3: Security enforcement and accountability.}
Security-relevant skill invocations should be authorized, bound to approved parameters, checked for freshness, protected against replay, and recorded with enough evidence to support audit, verification, and traceability.

\textbf{G4: Trusted or isolated enforcement support.}
Critical enforcement points should be protected by stronger isolation mechanisms when available, such as TEEs or microcontroller-level isolation.

\textbf{G5: Degraded execution under uncertainty.}
When the normal SHCUA-UAV chain cannot safely continue due to timeout, network delay, state mismatch, authorization failure, or environmental uncertainty, the onboard runtime should switch to bounded fallback behavior and preserve evidence for later analysis.

\section{Operation Modeling for SHCUA-UAV Control}

This section introduces the operation model that bridges SHCUA-style tool use and UAV control.
The goal of the model is not merely to rename tool calls as UAV skills.
Rather, it defines when an SHCUA-generated decision is eligible to become a UAV-executable action.
The model captures three requirements that are absent or weak in ordinary SHCUA tool use: temporal validity, state-dependent executability, and security accountability.

The core idea is to transform free-form SHCUA tool invocation into contract-bound vehicle skill invocation.
A skill invocation is not treated as an immediate flight command.
It is a structured execution object with explicit timing, state, authorization, fallback, and evidence semantics.
This object is the common basis for mission compilation, onboard dispatch, security enforcement, trusted evidence generation, and degraded execution.

\subsection{From Host Operations to Vehicle Operations}

A conventional SHCUA host operation can be abstracted as a tool invocation over digital resources:
\[
\mathcal{O}_{host} = \langle tool, args, obs \rangle ,
\]
where \(tool\) denotes a host-side tool, \(args\) denotes the arguments passed to that tool, and \(obs\) denotes the observation returned to the agent.

For example, \(tool\) may be \texttt{open\_browser}, \texttt{edit\_file}, \texttt{run\_command}, or \texttt{send\_message}.
The corresponding \(args\) may be a URL, a file path, a shell command, or a message body.
Such operations are usually discrete and software-defined.
Although they may be security-sensitive, they normally do not have to track continuously evolving physical dynamics.

This abstraction is insufficient for UAV control.
A UAV operation affects a physical vehicle whose state may change while the SHCUA is reasoning, while a network request is delayed, or while an onboard component is checking the command.
A UAV operation must therefore combine the requested action with timing, state, authority, fallback, and evidence context.

We abstract a UAV operation as:
\[
\mathcal{O}_{uav} =
\langle skill, args, S_t, T, A, C, F, E \rangle .
\]

The elements are defined as follows.

\begin{itemize}
    \item \(skill\) denotes a bounded UAV capability exposed by the vehicle-side skill library, such as \texttt{takeoff}, \texttt{hover}, \texttt{inspect\_area}, \texttt{return\_home}, or \texttt{emergency\_land}.
    It is not a raw actuator command or an arbitrary program generated by the SHCUA.

    \item \(args\) denotes skill parameters.
    Examples include target region, waypoint, altitude, sensing mode, and duration.
    For instance, \texttt{inspect\_area} may take \(args=\{area=A, altitude=20m, pattern=grid\}\).

    \item \(S_t\) denotes the vehicle and environment state at time \(t\).
    It may include pose, battery level, GPS status, obstacle distance, communication quality, and mission phase.

    \item \(T\) denotes temporal metadata.
    We define \(T=\langle t_g, t_r, ddl, vw, cls \rangle\), where \(t_g\) is the generation time of the decision, \(t_r\) is the time at which it is received by the onboard runtime, \(ddl\) is the execution deadline, \(vw\) is the validity window, and \(cls\) is the timing class of the operation.
    The timing class may be \(\mathsf{semantic}\), \(\mathsf{tactical}\), or \(\mathsf{reflex}\).
    Semantic decisions can tolerate seconds or longer, tactical decisions require sub-second to second-level response, and reflex decisions should be handled by onboard controllers without waiting for SHCUA reasoning.

    \item \(A\) denotes the authority context.
    It includes the user role, mission token, permitted skill set, scope, and policy version.
    This field determines whether an SHCUA decision is authorized to request a given UAV action.

    \item \(C\) denotes safety and security constraints.
    Safety constraints include flight-envelope, geofence, obstacle-clearance, and battery limits.
    Security constraints include command integrity, anti-replay requirements, and binding between approved intent and executable parameters.

    \item \(F\) denotes fallback behavior.
    It specifies what the onboard runtime should do if the invocation expires, becomes state-inconsistent, violates policy, or cannot be safely executed.
    Typical fallback actions include \texttt{hover}, \texttt{return\_home}, or \texttt{emergency\_land}.

    \item \(E\) denotes evidence requirements.
    It specifies the evidence needed for audit, including the user intent, compiled invocation, runtime decision, and final UAV behavior.
\end{itemize}

This abstraction captures the essential difference between host-side tool use and vehicle-side execution.
A host operation is mainly an invocation over software resources.
A UAV operation is a time-sensitive, state-dependent, authority-bound, and evidence-bearing action over a physical system.
Therefore, the SHCUA is no longer treated as a direct tool caller.
It becomes a task-level proposer whose intended operations must be transformed into vehicle operations with explicit timing, authority, and evidence semantics.

\subsection{Contract-Bound Skill Invocation}\label{subsec:contract-bound-skill-invocation}

A UAV skill invocation is the concrete execution object passed from the mission compiler to the onboard runtime.
We model an invocation as:
\[
\mathcal{I} =
\langle sid, src, op, args, pre, ddl, vw, sc, auth, inv, f_{\mathrm{b}}, ev \rangle .
\]

Here, \(sid\) is the skill invocation identifier.
The field \(src\) denotes the decision source, for example an offboard SHCUA, onboard runtime, or reflex controller.
The field \(op\) is the skill type, and \(args\) are the skill parameters.
The field \(pre\) denotes preconditions, \(ddl\) is the execution deadline, \(vw\) is the validity window, \(sc\) denotes state constraints, \(auth\) is the authority scope, \(inv\) denotes safety and security invariants, \(f_{\mathrm{b}}\) is the fallback action, and \(ev\) specifies evidence requirements.

The \(src\) field is important because not all decisions should be treated equally.
A cloud SHCUA may generate complex mission-level plans, but its decisions have longer and less predictable latency.
An onboard runtime or reflex controller has weaker semantic reasoning capability, but it can react quickly to local state changes.
The invocation model therefore does not assume that all decisions come from the same agent.
Instead, it records where the decision comes from and evaluates whether that decision source is appropriate for the required timing class.

For example, an instruction such as ``inspect area A at low altitude'' should not be compiled into a raw call such as \texttt{fly\_to(A)}.
Instead, it should produce a bounded invocation:
\[
\begin{aligned}
op &= \texttt{inspect\_area},\\
args &= \{area=A, altitude=20m, pattern=\texttt{grid}\},\\
pre &= \{\text{GPS available}, \text{battery} > 35\%\},\\
ddl &= \text{start within 2s},\\
vw &= \text{valid only in current mission phase},\\
sc &= \{\text{communication available}, \text{wind below threshold}\},\\
auth &= \{\text{inspection-only}, \text{area}=A, \text{altitude}\leq 30m\},\\
inv &= \{\text{no-fly zones excluded}, \text{no person tracking}\},\\
f_{\mathrm{b}} &= \{\texttt{hover}, \texttt{return\_home}\},\\
ev &= \{\text{user intent}, \text{agent decision}, \\
   &\quad \text{state snapshot}, \text{authorization result}\}.
\end{aligned}
\]

This representation is not merely an API wrapper.
It is the mechanism by which the SHCUA's open-ended reasoning is constrained before it can affect UAV behavior.
The SHCUA may propose a mission-level decision, but only the compiled invocation can enter the onboard execution path.

\subsection{Temporal and State Constraints}\label{sec:operation-model}

The real-time executability of a UAV skill invocation depends on both timing and state consistency.
An SHCUA-generated command may be semantically meaningful at generation time, but it may no longer be executable at dispatch time if the vehicle state, environment state, or mission phase has changed.
Thus, the deadline \(ddl\) is necessary but not sufficient.
It specifies the latest acceptable dispatch time, while the validity window, current state constraints, decision source, timing class, and remaining execution slack jointly determine whether the invocation can still enter the UAV execution path.

We distinguish three timing classes for SHCUA-UAV decisions, according to the role of the decision and the latency it can tolerate.
\begin{enumerate}
    \item \textbf{Semantic decisions.}
    A \(\mathsf{semantic}\) decision concerns mission-level intent, such as choosing an inspection order, selecting target regions, or generating a long-horizon mission plan.
    This class answers what the UAV should accomplish and how the mission should be organized.
    Such decisions may be produced by a cloud or edge SHCUA because they rely on semantic understanding and do not directly participate in the fast control loop.

    \item \textbf{Tactical decisions.}
    A \(\mathsf{tactical}\) decision concerns local skill selection or mission adjustment, such as skipping an unsafe waypoint or selecting the next executable skill.
    This class answers which bounded skill should be executed next under the current vehicle and environment state.
    Such decisions require bounded latency and should be handled by an onboard runtime, an onboard lightweight model, or a nearby edge component.

    \item \textbf{Reflex decisions.}
    A \(\mathsf{reflex}\) decision concerns immediate vehicle protection, such as collision avoidance, link-loss handling, or emergency landing.
    This class answers how the vehicle should react immediately to avoid unsafe behavior.
    Such decisions must not wait for SHCUA reasoning and should be handled by onboard controllers, safety supervisors, or the flight-control stack.
\end{enumerate}

The field \(ddl\) of invocation $\mathcal{I}$ denotes the latest time by which an invocation must be dispatched or started.
It provides a hard upper bound for execution.
The validity window \(vw\), in contrast, describes the time interval and mission context in which the decision remains applicable.
For example, a command to inspect an area may remain valid during the current mission phase, whereas a command to avoid a nearby obstacle may become invalid within a very short time because the relative position between the UAV and the obstacle changes rapidly.
Therefore, an invocation may be rejected even if it has not missed its deadline, as long as its state assumptions or mission context no longer hold.

We define temporal and state validity as:
\[
\begin{aligned}
valid(\mathcal{I}, S_t, t)
&=
valid_T(\mathcal{I}, t)
\wedge
valid_S(\mathcal{I}, S_t),
\end{aligned}
\]
where \(valid_T(\mathcal{I}, t)\) checks whether the current time \(t\) falls within the invocation's temporal constraints, including \(ddl\) and \(vw\), and \(valid_S(\mathcal{I}, S_t)\) checks whether the current UAV and environment state satisfies the invocation's state constraints.

State constraints may cover battery, GPS, velocity, obstacle distance, communication quality, mission phase, and local uncertainty.
For instance, a \texttt{track\_target} invocation may require sufficient battery, a valid perception result, a permitted target class, and a safe separation distance.
If any of these state assumptions no longer holds, the invocation should not be dispatched even if it is still within its specified deadline.

To model the tradeoff between slow SHCUA reasoning and fast UAV execution, we use the remaining latency before dispatch.
Let \(Q_{onboard}\) denote the onboard invocation queue.
If an invocation has already been generated, compiled, and placed in \(Q_{onboard}\), then the remaining latency only includes runtime validation and dispatch:
\[
L_{rem}(\mathcal{I}) = L_{val} + L_{disp},
\quad \mathcal{I} \in Q_{onboard}.
\]
If the invocation still depends on an offboard SHCUA decision or mission compilation, the remaining latency also includes source-side reasoning and compilation:
\[
L_{rem}(\mathcal{I}) =
L_{src} + L_{comp} + L_{val} + L_{disp},
\quad \mathcal{I} \notin Q_{onboard}.
\]
Here, \(L_{src}\) is the latency of the decision source, \(L_{comp}\) is the mission compilation latency, \(L_{val}\) is the runtime validation latency, and \(L_{disp}\) is the skill dispatch latency.

An invocation is admissible for execution only if it is temporally valid, state-consistent, and still has enough remaining slack before its deadline:
\[
\begin{aligned}
admissible(\mathcal{I}, S_t, t)
&=
valid(\mathcal{I}, S_t, t) \\
&\quad \wedge
\left(
L_{rem}(\mathcal{I})
\leq
ddl(\mathcal{I}) - t
\right).
\end{aligned}
\]

This admissibility condition explains how the model balances slow semantic reasoning and fast UAV execution.
For semantic tasks, the SHCUA can reason offboard and the resulting invocations can be compiled ahead of execution and queued onboard.
At dispatch time, the UAV does not wait for the SHCUA; it only performs validation and dispatch.
For tactical tasks, the invocation is admissible only if the remaining source, compilation, validation, and dispatch latency can still fit within the deadline.
For reflex-level tasks, SHCUA-generated invocations will normally not satisfy the admissibility condition because their remaining latency is too large for immediate vehicle protection.
Such actions must therefore be handled by onboard controllers, safety supervisors, or the flight-control stack.

If an invocation is not admissible, the onboard runtime must not simply wait for the SHCUA to finish reasoning.
It must reject the invocation, request replanning, or trigger a bounded fallback action such as \texttt{hover}, \texttt{return\_home}, or \texttt{emergency\_land}.
In this way, slow SHCUA reasoning can still contribute to mission-level intelligence, but it cannot block or override the fast execution path required by UAV control.

\subsection{Authorization and Evidence Semantics}

Security enforcement is modeled as a decision over the invocation, current state, and policy context:
\[
\eta(\mathcal{I}, S_t, P) \rightarrow
\{\mathsf{allow}, \mathsf{deny}, \mathsf{defer}, \mathsf{fallback}\}.
\]

Here, \(S_t\) is the current UAV state and \(P\) is the policy and authority context.
The result \(\mathsf{allow}\) permits execution, \(\mathsf{deny}\) rejects the invocation, \(\mathsf{defer}\) requests user confirmation or replanning, and \(\mathsf{fallback}\) triggers bounded degraded execution.

Authorization is not limited to checking whether a skill name is allowed.
It must bind the approved user intent, skill type, parameters, mission phase, time window, and authority scope.
For example, a user may be authorized to inspect a bridge from a safe distance, but not to track a person, enter a restricted area, descend below a minimum altitude, or continue the mission after the authorized time window expires.
The authorization result is therefore bound to the specific invocation:
\[
B = H(sid, op, args, auth, ddl, vw, P_{ver}),
\]
where \(H(\cdot)\) denotes a collision-resistant hash or authenticated binding function and \(P_{ver}\) denotes the policy version.
Any later modification of the operation, parameters, authorization scope, deadline, validity window, or policy version invalidates the binding.

Evidence semantics specify what must be preserved for accountability.
For each security-relevant invocation, the system records:
\[
ev =
\langle
h_{intent}, h_{agent}, \mathcal{I}, S_t, P_{ver}, r_{\eta}, r_{time}, B, f_{\mathrm{b}}, act
\rangle ,
\]
where \(h_{intent}\) is the hash of the user intent, \(h_{agent}\) is the hash of the SHCUA decision, \(P_{ver}\) is the policy version, \(r_{\eta}\) is the authorization result, \(r_{time}\) is the deadline and validity result, \(B\) is the command binding, \(f_{\mathrm{b}}\) is the fallback decision if any, and \(act\) is the final UAV behavior.

This evidence chain supports audit, verification, and traceability.
It links the original user intent, the SHCUA decision, the compiled invocation, the runtime authorization result, the state under which the decision was made, and the final vehicle behavior.

When trusted or isolated enforcement is available, the authorization decision, command binding, freshness checking, and evidence generation can be placed inside a protected boundary.
This protected boundary may be a TEE on an onboard or edge platform, or MPU/PMP/TrustZone-M-like isolation on a resource-constrained flight controller.
The full SHCUA and the high-frequency flight-control loop do not need to be moved into the trusted component.
Only the lower-frequency but security-critical control points are protected.

\subsection{Formal Properties}

We now state the formal properties provided by the proposed operation model and runtime enforcement.
These properties characterize how SHCUA-generated decisions are admitted into the UAV execution path.
They show that an invocation can affect the UAV only through mediated dispatch, and only when it satisfies timing, state, authorization, freshness, and evidence requirements.

We use \(valid(\mathcal{I}, S_t, t)\), \(admissible(\mathcal{I}, S_t, t)\), \(\eta(\mathcal{I}, S_t, P)\), the binding value \(B\), and the evidence tuple \(ev\) as introduced above.
The following assumptions make explicit the enforcement conditions required by the architecture.

\noindent\textbf{Assumption 1: Mediated dispatch.}
Every SHCUA-generated UAV action must be compiled into a skill invocation \(\mathcal{I}\) and submitted to the onboard runtime.
The SHCUA runtime has no direct channel to the UAV skill executor or the flight controller.

\noindent\textbf{Assumption 2: Pre-dispatch enforcement.}
Before an invocation reaches the UAV skill executor, the onboard runtime enforces admissibility, authorization, freshness, and command-binding checks.
An invocation is dispatched only if these checks succeed.

\noindent\textbf{Assumption 3: Bounded fallback.}
If an invocation is not admissible or fails enforcement, the onboard runtime either rejects it, requests replanning or confirmation, or triggers a bounded fallback action \(f_{\mathrm{b}}\), such as \texttt{hover}, \texttt{return\_home}, or \texttt{emergency\_land}.

\noindent\textbf{Assumption 4: Protected enforcement path.}
For security-critical invocations, authorization, command binding, freshness checking, and evidence generation are performed in a trusted or isolated enforcement path that cannot be bypassed by the ordinary SHCUA runtime.

\noindent\textbf{Theorem 1: Real-time admission control.}
Under Assumptions 1--3, an SHCUA-generated invocation that is expired, state-inconsistent, or lacks sufficient remaining execution slack cannot be dispatched to the UAV skill executor.

\noindent\textbf{Proof.}
By Assumption 1, any SHCUA-generated UAV action must first enter the onboard runtime as an invocation \(\mathcal{I}\).
By Assumption 2, the runtime dispatches \(\mathcal{I}\) only if the admissibility check succeeds.
According to the admissibility condition defined above, this requires both \(valid(\mathcal{I}, S_t, t)\) and
\[
L_{rem}(\mathcal{I}) \leq ddl(\mathcal{I}) - t .
\]
The validity predicate further requires that both temporal validity and state validity hold.
Therefore, if \(\mathcal{I}\) is expired, inconsistent with the current state \(S_t\), or cannot be dispatched within the remaining slack before its deadline, then \(admissible(\mathcal{I}, S_t, t)\) is false.
By Assumption 3, the runtime must reject the invocation, request replanning or confirmation, or trigger fallback instead of dispatching it.
Thus, such an invocation cannot reach the UAV skill executor.
\(\square\)

\noindent\textbf{Corollary 1: Non-blocking SHCUA reasoning.}
Slow SHCUA reasoning cannot force the UAV to wait indefinitely on a time-critical execution path.

\noindent\textbf{Proof.}
If the SHCUA or the communication path is slow, the remaining latency \(L_{rem}(\mathcal{I})\) increases.
Once \(L_{rem}(\mathcal{I}) > ddl(\mathcal{I}) - t\), the invocation becomes inadmissible and, by Theorem~1, cannot be dispatched.
The runtime must then reject it, request replanning or confirmation, or trigger bounded fallback.
Therefore, SHCUA reasoning may contribute to later semantic or tactical decisions, but it cannot hold the UAV execution path indefinitely.
\(\square\)

\noindent\textbf{Theorem 2: Security admission control.}
Under Assumptions 1, 2, and 4, an unauthorized, tampered, expired, or replayed SHCUA-generated invocation cannot be dispatched as a valid UAV action without detection.

\noindent\textbf{Proof.}
By Assumption 1, every SHCUA-generated UAV action must pass through the onboard runtime and enforcement path.
By Assumption 2, dispatch requires authorization, freshness, and command-binding checks to succeed.

For an unauthorized invocation, the enforcement result satisfies \(\eta(\mathcal{I}, S_t, P)\neq allow\).
Therefore, the authorization check fails and the invocation cannot be dispatched.

For a tampered invocation, at least one bound field in the approved operation, timing, authority, or policy context differs from the authorized value.
Since the binding value
\[
B = H(sid, op, args, auth, ddl, vw, P_{ver})
\]
depends on these fields, and \(H(\cdot)\) is assumed to be collision-resistant or authenticated, the modified invocation fails binding verification except with negligible probability.

For an expired invocation, temporal validity fails and hence admissibility fails.
For a replayed invocation, the freshness check detects that the invocation identifier, nonce, or sequence number has already been consumed, or that the invocation is outside its validity window.

By Assumption 4, these checks cannot be bypassed by the ordinary SHCUA runtime.
Thus, an unauthorized, tampered, expired, or replayed invocation cannot be dispatched as a valid UAV action without detection.
\(\square\)

\noindent\textbf{Theorem 3: Evidence-preserving mediation.}
Under Assumptions 1, 2, and 4, every security-relevant invocation that is allowed, denied, deferred, or redirected to fallback is associated with an evidence record.

\noindent\textbf{Proof.}
By Assumption 1, every SHCUA-generated UAV action must enter the mediated execution path as an invocation.
By Assumption 2, the invocation is checked before dispatch, and the enforcement result must be one of \(\mathsf{allow}\), \(\mathsf{deny}\), \(\mathsf{defer}\), or \(\mathsf{fallback}\).
For each result, the enforcement path generates an evidence tuple:
\[
\begin{aligned}
ev = \langle
&h_{intent}, h_{agent}, \mathcal{I}, S_t, P_{ver}, \\
&r_{\eta}, r_{time}, B, f_{\mathrm{b}}, act
\rangle .
\end{aligned}
\]
If the invocation is allowed, \(act\) records the dispatched UAV action.
If it is denied or deferred, \(r_{\eta}\) records the enforcement result.
If fallback is triggered, \(f_{\mathrm{b}}\) records the selected fallback action and \(act\) records the resulting UAV behavior.
By Assumption 4, evidence generation is performed in a trusted or isolated path and cannot be silently suppressed or forged by the ordinary SHCUA runtime.
Therefore, every security-relevant outcome has a corresponding evidence record.
\(\square\)

These properties show how the operation model supports the central design goal of RT-SHCUA.
SHCUA decisions may guide UAV missions, but they cannot directly bypass timing, state, authorization, freshness, or evidence requirements before affecting the vehicle.

\section{SHCUA-UAV Architecture}

This section presents the RT-SHCUA architecture built around the operation model introduced in the previous section.
The architecture maps the three timing classes of SHCUA-UAV decisions to different execution paths.
Semantic decisions are handled by the SHCUA and mission compiler, tactical decisions are handled by the onboard runtime, and reflex decisions are handled by onboard safety mechanisms and the flight-control stack.
This organization allows natural-language intelligence to guide UAV missions while keeping time-critical execution close to the vehicle.

RT-SHCUA consists of four logical layers.
The semantic agent layer interprets user instructions and produces mission-level decisions.
The mission compilation layer converts these decisions into contract-bound skill invocations.
The onboard real-time execution layer maintains vehicle state, checks invocation admissibility, and dispatches bounded UAV skills.
The security and safety enforcement layer mediates authorization, command binding, freshness checking, safety validation, evidence generation, and degraded recovery.
Together, these layers form a mediated control path from natural-language task intent to UAV execution.

\subsection{Overview}

\begin{figure*}[!t]
    \centering
    \includegraphics[width=0.85\textwidth]{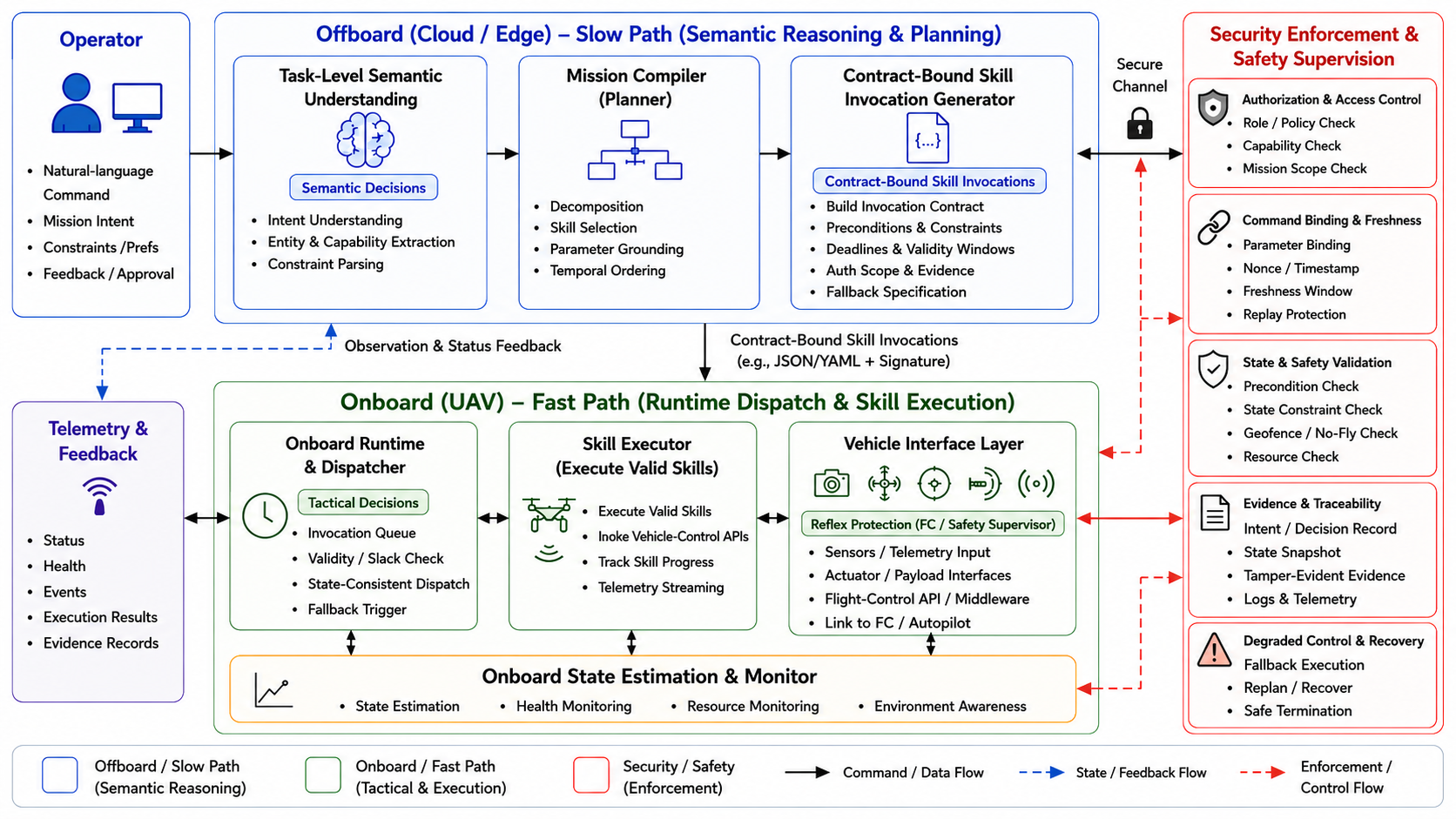}
    \caption{Overall architecture of the proposed RT-SHCUA framework for UAV control. The architecture separates offboard semantic reasoning from onboard tactical execution and reflex protection, and mediates contract-bound UAV skill invocations through security enforcement, safety supervision, evidence preservation, and degraded recovery mechanisms.}
    \label{fig:rt-shcua-arch}
\end{figure*}

Fig.~\ref{fig:rt-shcua-arch} illustrates the overall architecture.
The figure is organized around three execution paths and one feedback path.
The blue region at the top represents the offboard slow path, where the SHCUA interprets user intent, generates task-level mission decisions, and invokes the mission compiler.
The mission compiler converts these SHCUA decisions into contract-bound UAV skill invocations with explicit timing, state, authorization, fallback, and evidence requirements.
This path may use cloud or edge computing resources because SHCUA reasoning mainly concerns mission-level semantics rather than immediate vehicle protection.

The green region at the bottom represents the onboard fast path implemented above the existing flight-control stack.
It includes the onboard runtime, invocation queue, skill executor, and vehicle-control interface.
Through this interface, admissible skill invocations are translated into bounded UAV actions and submitted to the flight controller or autopilot.
The flight controller itself remains responsible for low-level stabilization, actuator control, and reflex-level protection, and provides vehicle state and execution feedback to the onboard runtime.

The red region on the right represents the security enforcement and safety supervision path.
It acts as the control boundary between SHCUA-generated decisions and UAV execution.
For each security-relevant invocation, this path verifies authorization, command binding, freshness, policy compliance, safety constraints, and evidence requirements.
When an invocation fails these checks, the enforcement layer rejects the invocation, requests confirmation or replanning, or triggers a bounded fallback action.
The dashed feedback path carries telemetry, execution status, events, enforcement results, fallback records, and evidence logs back to the semantic layer.
This feedback supports later replanning and audit without placing the SHCUA in the fast execution loop.

In RT-SHCUA, the SHCUA serves as the semantic entry point of the control chain.
It interprets user instructions, extracts mission intent, and produces task-level decisions.
These decisions are compiled into contract-bound skill invocations before entering the onboard execution path.
Each invocation then passes through admission, security/safety validation, and evidence generation before it can be dispatched to the UAV skill executor.
This mediated structure keeps time-critical execution on the onboard side, while allowing the SHCUA to contribute high-level mission intelligence through bounded and auditable skill invocations.

\subsection{Task-Level Semantic Understanding}

The task-level semantic understanding layer processes natural-language mission instructions.
It extracts the mission intent, operating constraints, target references, and approval requirements.
For example, a user may instruct the UAV to inspect a bridge while respecting safety boundaries and return conditions.
Such instructions require semantic interpretation and task decomposition before they can be mapped to UAV capabilities.

This layer may run on a cloud server, an edge node, or a capable onboard companion computer.
A large model can resolve ambiguous instructions, infer missing task context, and generate an ordered mission plan.
The output of this layer belongs mainly to the \(\mathsf{semantic}\) timing class.
It specifies the mission objective and organization, such as inspection order or termination conditions.

The semantic layer operates above the real-time vehicle execution path.
It produces mission descriptions, task plans, or structured intent representations.
The following layers convert these semantic outputs into bounded UAV skill invocations and decide when such invocations can be executed.
This preserves the flexibility of SHCUA-based natural-language interaction while keeping vehicle-level execution under onboard timing and enforcement constraints.

\subsection{Mission Compilation}

The mission compiler translates the semantic agent's output into a structured mission representation, such as a mission graph or skill sequence.
Each node in this representation is mapped to a bounded UAV skill and then instantiated as a contract-bound skill invocation.
The compiler is therefore the interface between open-ended language reasoning and vehicle-side execution.

For each invocation, the compiler attaches the fields defined in the operation model.
These fields capture the requested skill, parameters, timing/state constraints, authority scope, fallback behavior, and evidence requirements.
The compiler also assigns a timing class to each operation.
Mission-level steps are compiled as \(\mathsf{semantic}\) or \(\mathsf{tactical}\) invocations, while immediate protection behaviors are mapped to onboard reflex policies and fallback mechanisms.

This compilation step gives SHCUA-generated decisions an executable structure.
A natural-language plan may be underspecified, delayed, overly broad, or influenced by unsafe context.
The compiler resolves the plan into explicit runtime objects whose admissibility and enforcement conditions can be checked before execution.
In this way, UAV capabilities are exposed to the SHCUA through a controlled invocation interface rather than through raw flight commands or arbitrary code.

\subsection{Onboard Real-Time Skill Execution}

The onboard runtime is responsible for tactical execution.
It receives compiled invocations, maintains the invocation queue, tracks vehicle state, evaluates admission conditions, and dispatches admissible invocations to the skill executor.
The runtime operates close to the vehicle state and is designed for bounded-latency decisions.

The onboard runtime may contain a rule-based scheduler, a lightweight onboard model, or a hybrid decision module.
It handles local tactical decisions such as selecting the next executable skill, skipping an unsafe waypoint, or triggering fallback when an invocation becomes inadmissible.
These decisions correspond to the \(\mathsf{tactical}\) timing class and require faster response than offboard semantic planning.

The skill executor runs bounded UAV skills exposed by the vehicle-side skill library.
Representative skills include \texttt{takeoff}, \texttt{hover}, \texttt{inspect\_area}, \texttt{return\_home}, and \texttt{emergency\_land}.
Each skill is implemented against existing UAV control interfaces, such as a flight controller, autopilot, or middleware stack.
The skill executor tracks progress, reports execution results, and streams telemetry back to the runtime and semantic layer.

This layer provides the main real-time decoupling mechanism.
A high-level SHCUA plan may be generated slowly, but once the plan has been compiled and queued onboard, dispatch depends on local validation, enforcement, and skill execution rather than continuous offboard reasoning.
If an invocation is expired, state-inconsistent, unauthorized, or lacks sufficient slack, the runtime follows the contract by rejecting, deferring, or triggering fallback.

\subsection{Security Enforcement and Safety Supervision}

The enforcement layer mediates contract-bound invocations before they affect the UAV.
It contains two coordinated functions: security enforcement and safety supervision.

Security enforcement protects control authority, command integrity, freshness, and accountability.
It checks whether an invocation is authorized, fresh, bound to the approved mission scope, and accompanied by the required evidence.
This prevents an ordinary SHCUA runtime or compromised software component from expanding the task, replaying stale decisions, or producing untraceable UAV actions.

Safety supervision checks physical feasibility and accident-avoidance constraints.
It evaluates the current vehicle state against geofence, flight-envelope, battery, obstacle-clearance, and emergency constraints.
A command that is authorized at the security level may still be rejected by the safety supervisor if the current vehicle or environment state makes it physically unacceptable.

Security and safety address different parts of the UAV control problem.
Security enforcement determines whether an SHCUA-generated invocation has the authority and integrity required to enter the execution path.
Safety supervision determines whether the requested behavior remains physically acceptable under the current state.
RT-SHCUA requires both forms of mediation before a security-relevant invocation can affect the vehicle.

Security-critical enforcement points can be protected by trusted or isolated mechanisms.
On a resource-rich onboard or edge platform, a TEE can protect authorization, command binding, key material, policy versions, freshness checking, and evidence generation.
On a flight controller or resource-constrained MCU, MPU/PMP/TrustZone-M-like isolation can protect command-entry points, security monitor tasks, or fallback-triggering logic.
These mechanisms protect the lower-frequency but security-critical control points through which SHCUA decisions enter UAV execution.

\subsection{Degraded Execution and Recovery}

Degraded execution is the bounded recovery path used when the normal SHCUA-UAV execution chain cannot safely continue.
It is activated when an invocation becomes expired, state-inconsistent, unauthorized, or lacks sufficient remaining execution slack.
It can also be triggered by model delay, degraded sensing or communication, low battery, obstacle emergence, or low-confidence mission state.

Fallback actions include local hold, return-home, emergency landing, or user confirmation.
The fallback selection is deadline-aware and state-dependent.
For example, delayed semantic planning may trigger temporary hovering, low battery may trigger return-to-home, and a rapidly approaching obstacle may trigger immediate onboard avoidance or emergency landing.
The selected fallback action is bounded by policy and by the current UAV state.

Degraded execution provides continuity when the SHCUA cannot provide a timely or admissible decision.
The onboard runtime selects a fallback according to the invocation contract, the current state, and the enforcement result.
This prevents slow reasoning, unstable communication, or unsafe agent output from blocking the fast vehicle execution path.

Degraded execution also preserves accountability.
The system records the trigger condition, relevant invocation, state snapshot, enforcement result, fallback action, and final UAV behavior.
These records allow later audit and traceability for successful executions, rejected invocations, deferred decisions, and fallback-mediated behaviors.
Thus, RT-SHCUA maintains an evidence trail across both normal and degraded execution paths.

\section{Implementation}

This section presents the prototype implementation used to evaluate the feasibility and effectiveness of the proposed RT-SHCUA architecture and its key mechanisms.
The prototype is built on representative software artifacts at both ends of the SHCUA-to-UAV control chain:
OpenClaw provides the SHCUA-side agent execution model, while PX4 SITL with Gazebo provides the UAV execution backend and simulation environment.
This setting evaluates RT-SHCUA in a representative SHCUA-to-UAV control scenario grounded in existing agent and UAV software systems.

\begin{figure*}[!t]
\centering
\includegraphics[width=0.9\textwidth]{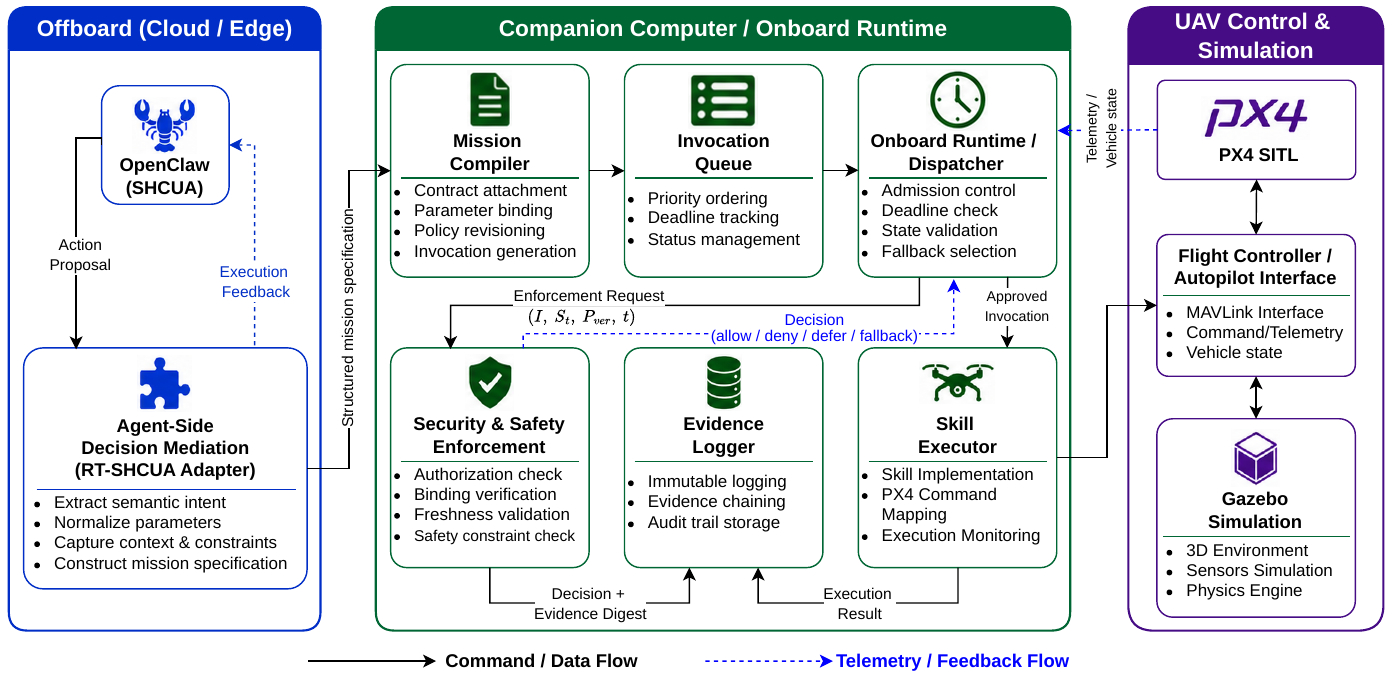}
\caption{Prototype implementation architecture of RT-SHCUA. OpenClaw provides the SHCUA-side agent execution model, PX4 SITL with Gazebo provides the UAV execution backend, and RT-SHCUA mediates between them by transforming agent-side action proposals into structured mission specifications, compiling them into contract-bound UAV skill invocations, enforcing runtime admission and security checks, and recording execution evidence.}
\label{fig:rt-shcua-imp}
\end{figure*}

RT-SHCUA is implemented as the mediation and enforcement layer between OpenClaw and the UAV execution backend.
As shown in Fig.~\ref{fig:rt-shcua-imp}, OpenClaw generates task-level action proposals from natural-language instructions.
The agent-side decision mediation component transforms these proposals into structured mission specifications.
The mission compiler converts the structured specifications into contract-bound UAV skill invocations.
The onboard runtime then admits, enforces, dispatches, and logs these invocations through runtime admission, security checking, fallback handling, and evidence preservation.
This implementation directly instantiates the main mechanisms of RT-SHCUA on top of existing SHCUA and UAV software systems.

\subsection{Agent-Side Decision Mediation}

The component labeled ``Agent-Side Decision Mediation'' in Fig.~\ref{fig:rt-shcua-imp} implements the input boundary between OpenClaw's task-level reasoning and RT-SHCUA's contract-based UAV execution.
OpenClaw generates action proposals through its SHCUA loop, which combines language understanding, tool-oriented reasoning, feedback, and replanning.
The mediation component transforms these action proposals into structured mission specifications for downstream compilation.

The structured mission specification captures the task objective, target region, requested UAV capability, execution parameters, and mission constraints.
For example, an OpenClaw proposal to inspect a target region is represented by the inspection target, altitude preference, sensing requirement, mission scope, and fallback preference.
This representation captures the semantic intent of the SHCUA decision and provides the input format required by the mission compiler.

This mediation step connects open-ended SHCUA reasoning with contract-based UAV execution.
SHCUA outputs are typically context-dependent and tool-oriented, while UAV execution requires bounded operations with explicit timing, state, and authority semantics.
The mediation component therefore provides a stable input boundary for RT-SHCUA.
Its output is a structured mission specification that can be compiled into contract-bound skill invocations and further processed by runtime admission, enforcement, fallback handling, and evidence preservation.

\subsection{Contract-Bound Mission Compilation}

The component labeled ``Mission Compiler'' in Fig.~\ref{fig:rt-shcua-imp} transforms structured mission specifications into contract-bound UAV skill invocations.
The invocation format follows the contract-bound skill invocation model defined in Section~\ref{subsec:contract-bound-skill-invocation}.
In the prototype, the mission compiler instantiates this model by assigning the invocation identifier, decision source, UAV operation, operation parameters, preconditions, dispatch deadline, validity window, state constraints, authorization scope, safety and security invariants, fallback behavior, and evidence requirements.
The resulting invocation is stored as an immutable runtime record and inserted into the invocation queue.

The prototype implements representative UAV skills, including \texttt{takeoff}, \texttt{inspect\_area}, \texttt{return\_home}, and \texttt{emergency\_land}.
Each skill is associated with a contract schema that defines its admissible parameter range, required vehicle state, temporal profile, authorization requirement, fallback policy, and evidence hooks.
For example, \texttt{inspect\_area} requires a valid target region, bounded altitude, sufficient battery, and a fallback action.
The schema turns each UAV capability into a checked execution primitive that can be processed by the runtime and enforcement components.

During compilation, the compiler also generates authentication metadata for each invocation.
Specifically, it computes a binding value over selected security-relevant fields:
\begin{equation*}
B = H(sid, src, op, args, ddl, vw, auth, inv, P_{ver}).
\end{equation*}

Here, \(P_{ver}\) denotes the policy version used during compilation.
The value \(B\), together with its signature or MAC, is stored as implementation metadata associated with the invocation.
This binding links the operation, parameters, timing constraints, authorization scope, security invariants, and policy version to the compiled invocation.
The enforcement component later uses this metadata to verify that the invocation being dispatched is consistent with the invocation produced by the mission compiler.

\subsection{Runtime Admission and Skill Dispatch}

The components labeled ``Invocation Queue'', ``Onboard Runtime/Dispatcher'', and ``Skill Executor'' in Fig.~\ref{fig:rt-shcua-imp} implement the runtime execution path for compiled invocations.
After mission compilation, contract-bound invocations are inserted into the invocation queue.
The onboard runtime maintains the queue, tracks mission phase, collects current UAV state, and evaluates whether each queued invocation remains admissible at dispatch time.

For each invocation, the runtime evaluates temporal validity, state consistency, remaining execution slack, and the enforcement result produced by the security and safety enforcement component.
This admission procedure instantiates the admissibility condition defined in Section~\ref{sec:operation-model}.
An invocation is dispatched when its validity window still holds, its state assumptions match the current UAV state, and its remaining execution latency fits within the available deadline slack.

The runtime produces one of four execution outcomes for each invocation.
If the invocation satisfies the admission and enforcement checks, the runtime dispatches it to the skill executor.
If the invocation violates temporal, state, or authorization constraints, the runtime rejects it and records the corresponding decision.
If the invocation requires additional state updates or confirmation, the runtime defers its execution.
If the invocation cannot proceed under the current mission state, the runtime triggers the fallback behavior specified in the invocation contract.

The skill executor maps admitted invocations to bounded UAV capabilities exposed by the UAV execution backend.
For example, skills such as \texttt{hover}, \texttt{inspect\_area}, and \texttt{return\_home} are invoked through the PX4 control interface.
During execution, the skill executor reports status and telemetry to the onboard runtime.
This runtime path ties SHCUA-generated task decisions to local vehicle state, timing constraints, and enforcement results before they affect UAV behavior.

\subsection{Security Enforcement and Evidence Preservation}

The components labeled ``Security \& Safety Enforcement'' and ``Evidence Logger'' in Fig.~\ref{fig:rt-shcua-imp} implement the security-critical checks and audit path for compiled invocations.
Before an admitted operation reaches the skill executor, the enforcement component evaluates authority, binding, freshness, policy version, and safety constraints.

Authorization checking evaluates whether the requested operation and mission context are permitted under the current policy.
Binding verification evaluates whether the invocation fields used during compilation remain consistent at dispatch time.
In the prototype, this check uses binding metadata generated by the mission compiler over the invocation identity, operation, timing, authority, invariants, and policy version.
Freshness checking uses identifiers, timestamps, nonces, and validity windows to maintain a consistent execution history.

The enforcement component returns an execution decision to the onboard runtime.
The decision belongs to one of four categories: \(\mathsf{allow}\), \(\mathsf{deny}\), \(\mathsf{defer}\), or \(\mathsf{fallback}\).
An \(\mathsf{allow}\) decision permits the runtime to dispatch the invocation.
A \(\mathsf{deny}\) decision records the rejected invocation and its enforcement reason.
A \(\mathsf{defer}\) decision keeps the invocation pending until the required state or confirmation becomes available.
A \(\mathsf{fallback}\) decision activates the recovery behavior specified by the invocation contract.

The evidence logger records security-relevant execution events along the mediated control path.
Each evidence record is linked to the corresponding invocation and captures the user intent reference, compiled invocation, policy version, state snapshot, enforcement result, fallback action, and final outcome.
Evidence records can be stored in an append-only log or organized as a hash chain.
This mechanism provides an auditable record for successful dispatches, rejected invocations, deferred decisions, and fallback-mediated executions.

\subsection{Trusted Enforcement Boundary}

The components named ``Security \& Safety Enforcement'' and ``Evidence Logger'' in Fig.~\ref{fig:rt-shcua-imp} contain the security-critical control points of RT-SHCUA.
These control points determine whether an SHCUA-derived invocation can enter UAV execution and whether the corresponding execution record can be preserved for later audit.
The protected functions include authorization, binding, freshness, key protection, and evidence-record generation.

\subsubsection{Boundary Placement}

RT-SHCUA places the trusted boundary around the admission and evidence path.
The onboard runtime prepares an enforcement request from the compiled invocation, current state snapshot, policy version, and timing context.
The enforcement component evaluates the request and returns an execution decision to the runtime.
The runtime then dispatches the invocation, rejects it, defers it, or triggers fallback according to the returned decision.

The trusted boundary covers the security-critical logic required by this decision path.
OpenClaw, mission compilation, invocation queuing, skill execution, PX4 SITL, Gazebo, and the flight-control stack remain in their normal execution domains.
This boundary placement keeps the protected computing base focused on the control points that mediate SHCUA-generated decisions before they affect UAV behavior.

\subsubsection{Software Enforcement Baseline}

The prototype first implements the enforcement component as a software baseline on the companion-computer side.
In this configuration, authorization, binding, freshness, and evidence logging are executed as ordinary runtime components.
The software baseline is used to validate the end-to-end execution logic and to provide a reference point for measuring the overhead introduced by trusted enforcement.

\subsubsection{ARM TrustZone-based Enforcement Prototype}

\begin{figure}[t]
\centering
\includegraphics[width=\linewidth]{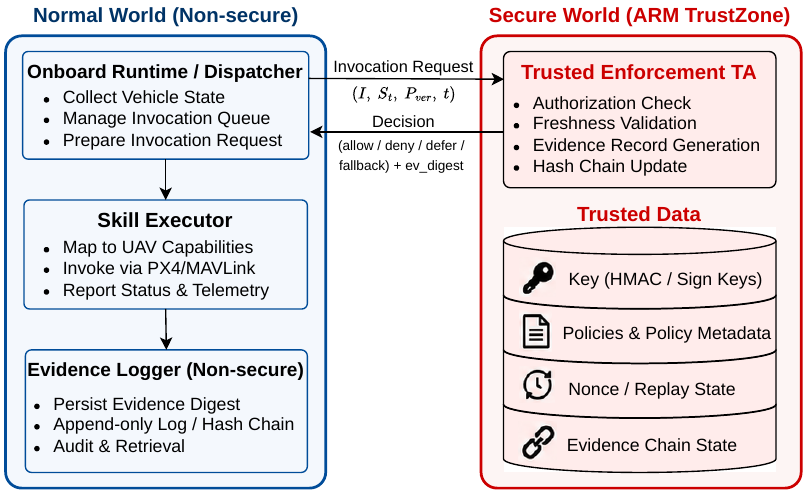}
\caption{ARM TrustZone-based implementation of the trusted enforcement boundary. The Normal World submits invocation requests to the Trusted Enforcement TA in the Secure World and receives enforcement decisions and evidence digests before dispatching UAV skills.}
\label{fig:trustzone-enforcement}
\end{figure}

To validate the feasibility of hardware-assisted enforcement, we instantiate the security-critical enforcement path using ARM TrustZone.
Fig.~\ref{fig:trustzone-enforcement} shows the TrustZone-based implementation.
The Normal World hosts the onboard runtime, skill executor, and non-secure evidence logger.
The Secure World hosts a Trusted Enforcement TA and the trusted state required for authorization, binding, freshness, and evidence-digest generation.

The onboard runtime invokes the Trusted Enforcement TA through a narrow interface:
\begin{equation*}
\mathsf{enforce}(\mathcal{I}, S_t, P_{ver}, t)
\rightarrow
\langle r, c, d_{ev} \rangle .
\end{equation*}

Here, \(\mathcal{I}\) is the compiled invocation, \(S_t\) is the current UAV state snapshot, \(P_{ver}\) is the policy version, and \(t\) is the admission timestamp.
The return value \(r \in \{\mathsf{allow}, \mathsf{deny}, \mathsf{defer}, \mathsf{fallback}\}\) is the enforcement decision, \(c\) is the reason code, and \(d_{ev}\) is the evidence digest generated inside the Secure World.

The Trusted Enforcement TA implements the security checks required by the RT-SHCUA admission path.
It verifies the authorization scope, command binding, freshness state, and policy version associated with the invocation.
It maintains protected data for keys, policy metadata, replay state, and the evidence chain.

For each enforcement request, the Trusted Enforcement TA binds the decision to the invocation identity, selected operation, policy version, timing result, and UAV state snapshot.
It then updates the protected evidence-chain state and returns the evidence digest \(d_{ev}\) to the Normal World.
The non-secure evidence logger persists the full evidence record together with \(d_{ev}\), enabling later verification that the stored record is consistent with the trusted enforcement result.

This implementation demonstrates that the security-critical part of invocation admission can be isolated with ARM TrustZone.
The trusted code is limited to enforcement checks, protected metadata, freshness state, and evidence-chain updates, while the SHCUA reasoning loop, mission compilation logic, skill execution logic, and flight-control backend remain in the normal execution path.

\section{Evaluation}

This section evaluates RT-SHCUA at the task-level mediation boundary. UAV
missions run in PX4 SITL/Gazebo, while the trusted admission path is measured
with a QEMU-based OP-TEE TrustZone emulation environment. The experiments focus
on four questions: whether normal dispatch remains bounded, whether longer
missions destabilize the admission path, whether degraded/security cases produce
correct bounded behavior, and whether trusted enforcement and evidence logging
add acceptable overhead.

\subsection{Experimental Platform and Configuration}

The evaluated prototype follows the implementation architecture shown in
Fig.~\ref{fig:rt-shcua-imp} and the protected enforcement boundary shown in
Fig.~\ref{fig:trustzone-enforcement}. Table~\ref{tab:exp-platform} summarizes
the hardware, software, trusted-enforcement environment, and evaluation matrix.

\begin{table}[!htbp]
\caption{Experimental platform configuration.}
\label{tab:exp-platform}
\centering
\scriptsize
\setlength{\tabcolsep}{3pt}
\renewcommand{\arraystretch}{1.08}
\begin{tabularx}{\linewidth}{@{}p{0.34\linewidth}X@{}}
\hline
Item & Configuration \\
\hline

\multicolumn{2}{@{}l}{\textit{Simulation Platform}} \\
Simulation server & Intel Xeon server \\
CPU & Intel Xeon Platinum 8558 (48 cores @ 2.1\,GHz) \\
Memory & 256\,GB DDR5 (4800\,MT/s) \\
Storage & Samsung SSD 990 EVO Plus (2\,TB) \\
\hline

\multicolumn{2}{@{}l}{\textit{Software Stack}} \\
Operating system & Ubuntu 24.04 LTS (Linux 6.8.x) \\
Build environment & GCC/G++ 13.x, CMake, Ninja, Python 3.12.x \\
Agent runtime & OpenClaw \\
RT-SHCUA layer & Agent-side adapter, mission compiler, onboard runtime, enforcement client, and evidence logger \\
Back-end LLM & Qwen-Max LLM (temperature = 0) \\
Flight-control stack & PX4 SITL (PX4-Autopilot v1.16.0) \\
Simulator & Gazebo Harmonic LTS \\
UAV interface & MAVSDK-Python 3.15.3 / MAVLink v2 \\
\hline

\multicolumn{2}{@{}l}{\textit{Trusted Enforcement}} \\
TEE environment & QEMU-based OP-TEE \\
Normal World & Linux guest running the enforcement client \\
Secure World & OP-TEE Trusted Enforcement TA \\
TA toolchain & OP-TEE cross-compilation toolchain \\
Protected state & Keys, policy metadata, replay nonce state, and evidence-chain state \\
Secure storage & OP-TEE secure storage \\
\hline

\multicolumn{2}{@{}l}{\textit{Evaluation Matrix}} \\
Mission templates & Short, standard, and extended \\
Task-level deadlines & 0.1, 0.2, 0.3, 0.5, 0.75, 1.0, and 1.5\,s \\
Enforcement modes & None, software, and OP-TEE-backed \\
Repeated runs & Thirty runs for each matrix cell \\
Normal matrix & $3 \times 7 \times 3 \times 30 = 1890$ executions \\
Security/degraded checks & $12 \times 21 \times 30 = 7560$ trusted-path checks \\
\hline

\end{tabularx}
\end{table}

The realtime matrix contains 63 mission--deadline--mode cells, with thirty
runs per cell. It therefore produces 1890 normal mission executions. Latency
values are aggregated over the repeated runs using median and tail-latency
statistics. The security/degraded suite reuses the same mission--deadline points
and evaluates 12 representative cases, resulting in 7560 trusted-path checks.
Dedicated fallback-latency and evidence-audit measurements are reported where
they are relevant. All task-level deadlines apply to the RT-SHCUA
skill-admission and dispatch boundary, not to low-level vehicle-control loops.

Table~\ref{tab:onboard-footprint} reports the footprint of the RT-SHCUA-added
mediation, enforcement, and evidence-preservation components. It excludes the
offboard SHCUA runtime, the back-end language model, PX4, Gazebo, MAVSDK/MAVLink
libraries, OP-TEE OS, and the existing flight-control stack.

\begin{table}[!htbp]
\caption{RT-SHCUA onboard footprint.}
\label{tab:onboard-footprint}
\centering
\scriptsize
\setlength{\tabcolsep}{1.8pt}
\renewcommand{\arraystretch}{1.08}
\begin{tabularx}{\linewidth}{@{}p{0.27\linewidth}p{0.09\linewidth}p{0.12\linewidth}p{0.15\linewidth}p{0.14\linewidth}X@{}}
\hline
Component & Domain & Eff. LoC & Artifact Size & Peak Mem. & State \\
\hline
Runtime dispatcher & CC & 134 & 8.29 KB & 4.76 KB & -- \\
Invocation queue & CC & 23 & 8.29 KB & 16.32 KB & Queue state \\
State monitor & CC & 102 & 12.38 KB & 10.05 KB & Vehicle-state snapshot \\
Skill executor & CC & 127 & 10.50 KB & 0.31 KB & -- \\
Enforcement client & CC & 215 & 11.80 KB & 12.06 KB & Policy cache \\
Evidence logger & CC & 53 & 2.63 KB & 184.24 KB & Evidence chain \\
Trusted enforcement TA & SW & 642 & 119.55 KB & 310.10 KB & Keys, nonces, chain \\
\hline
Total & -- & 1292 & 156.43 KB & 488.52 KB & -- \\
\hline
\multicolumn{6}{@{}p{\linewidth}@{}}{\scriptsize
CC denotes the companion-computer runtime domain; SW denotes the emulated
OP-TEE Secure World. Eff. LoC excludes blanks, comments, and docstrings.
Artifact sizes may include shared files; Total is de-duplicated. CC peak memory
uses Python \texttt{tracemalloc}, while TA peak memory is estimated from ELF
sections and configured stack/data sizes.}
\end{tabularx}
\end{table}

\subsection{Responsiveness and Mission Scaling}

\begin{figure*}[!htbp]
\centering
\includegraphics[width=0.9\linewidth]{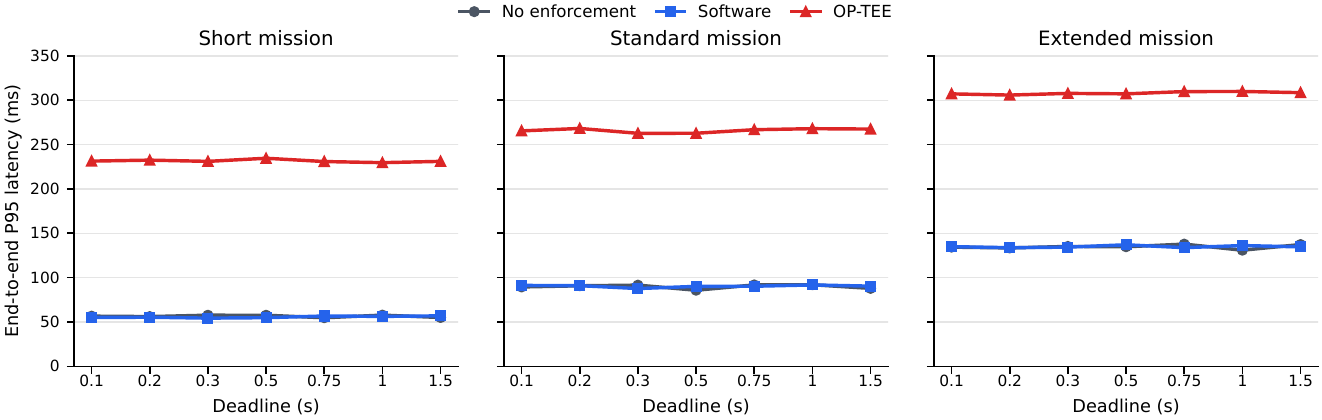}
\caption{Normal matrix P95 end-to-end latency under three enforcement modes,
three mission lengths, and 0.1--1.5\,s task-level deadlines. The matrix reports
zero task-level deadline-miss and fallback events across all 1890 executions.}
\label{fig:e2e-responsiveness}
\end{figure*}

We first evaluate whether RT-SHCUA keeps SHCUA-generated UAV skill invocations
within a bounded task-level dispatch path. Fig.~\ref{fig:e2e-responsiveness}
reports the cell-averaged P95 end-to-end latency across the normal matrix. The
no-enforcement and software-enforcement paths are almost identical: their
averaged P95 latencies are 56.5/55.7\,ms for short missions, 89.9/90.4\,ms for
standard missions, and 134.9/135.0\,ms for extended missions. Software
enforcement stays close to the no-enforcement baseline because it runs inside
the companion-computer admission path and adds only sub-millisecond validation
cost.

The OP-TEE-backed path raises the corresponding P95 latencies to 231.6\,ms,
265.9\,ms, and 308.0\,ms. The increase comes mainly from crossing the
normal-world/secure-world boundary and executing protected validation inside the
TA. Even with this added cost, the normal matrix reports zero task-level
deadline-miss flags and zero fallback events across all 1890 normal executions.
This indicates that the trusted path remains bounded at the skill-admission
boundary, although it is intended for lower-rate mission-level skills rather
than high-frequency vehicle-control loops.

\begin{figure}[!tbp]
\centering
\captionsetup[subfloat]{font=scriptsize}
\subfloat[Median end-to-end latency.]{%
\includegraphics[width=0.98\linewidth]{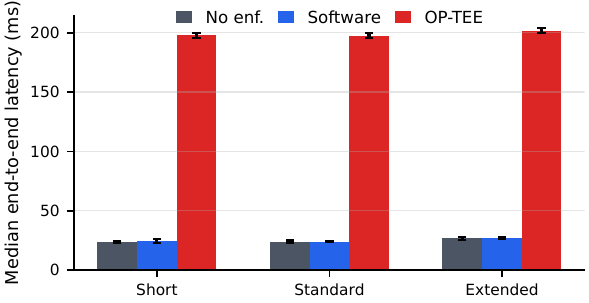}}
\par\vspace{0.25em}
\subfloat[P95 validation latency (log scale).]{%
\includegraphics[width=0.98\linewidth]{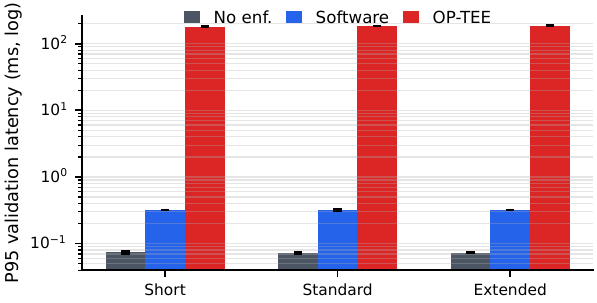}}
\caption{Mission-scaling latency results. The left panel shows median
end-to-end latency, and the right panel shows P95 validation latency. Bars
average the seven deadline points; whiskers show the min--max range.}
\label{fig:mission-scaling}
\end{figure}

Fig.~\ref{fig:mission-scaling} further examines how latency changes with mission
length. The left panel shows median end-to-end latency, and the right panel
shows P95 validation latency. For no enforcement and software enforcement,
median end-to-end latency remains low as missions grow: 24.0/24.4\,ms for
short missions, 24.0/24.0\,ms for standard missions, and 26.8/27.2\,ms for
extended missions. The OP-TEE-backed medians are higher, at 197.7\,ms,
197.5\,ms, and 201.3\,ms, but remain nearly flat across mission lengths.

The validation results explain this pattern. P95 validation latency is about
0.073--0.074\,ms without enforcement, 0.318--0.322\,ms with software
enforcement, and 180.7--185.4\,ms with OP-TEE-backed enforcement. Therefore,
mission growth mainly increases the number of bounded skill invocations rather
than the cost of each local admission, while the trusted configuration pays a
stable per-invocation boundary cost.

\subsection{Degraded Execution and Fallback Behavior}

The normal realtime matrix is intentionally composed of admissible executions,
so it cannot by itself demonstrate degraded behavior. We therefore evaluate a
separate suite covering representative runtime degradation and security-triggering
cases, such as expired slack, missing GPS, policy violations, and replay. A
normal allow case is included as a control.

\begin{figure}[!tbp]
\centering
\includegraphics[width=\linewidth]{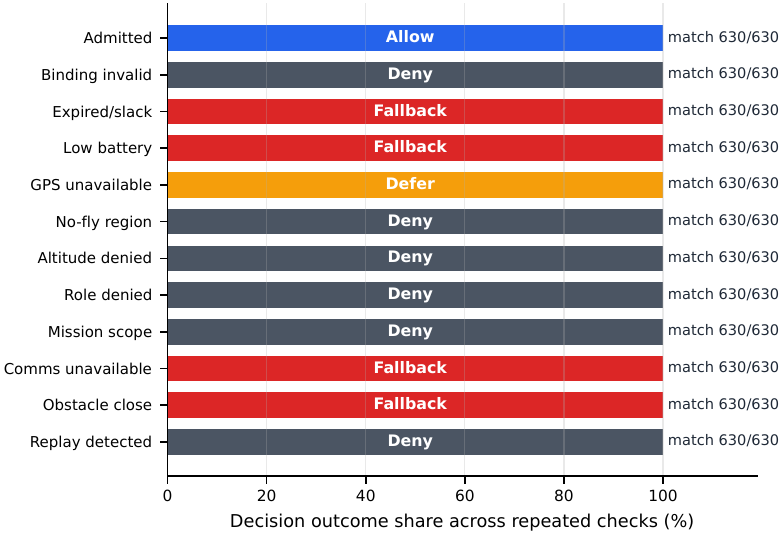}
\caption{Degraded and security-triggering decision outcomes. The OP-TEE-backed
trusted path matches the software reference for all repeated checks, while
recoverable degradation is mapped to fallback, insufficient state to defer, and
security violations to deny.}
\label{fig:degraded-execution}
\end{figure}

Fig.~\ref{fig:degraded-execution} reports the trusted-path decision outcomes.
Across 12 cases, 21 mission--deadline points, and thirty repetitions, the
OP-TEE-backed path matches the software reference in all 7560 checks. The
outcome distribution follows the intended semantics: recoverable runtime
degradation is converted into fallback, insufficient state is deferred, and
authority, scope, integrity, or freshness violations are denied. This separation
is important: fallback is a bounded safety-recovery mechanism, not a way to
execute unauthenticated or out-of-contract commands.

A dedicated fallback microbenchmark further measures local degraded handling
with 30 runs per triggered condition. The measurement covers local admission,
decision selection, and evidence generation. Median dispatch latencies are tightly
clustered between 0.132\,ms and 0.143\,ms, and P95 latencies remain below
0.149\,ms for the measured cases. Thus, degraded handling is a bounded local
runtime path rather than a slow replanning loop through the SHCUA.

\subsection{Trusted Enforcement Overhead}

\begin{figure}[!tbp]
\centering
\includegraphics[width=\linewidth]{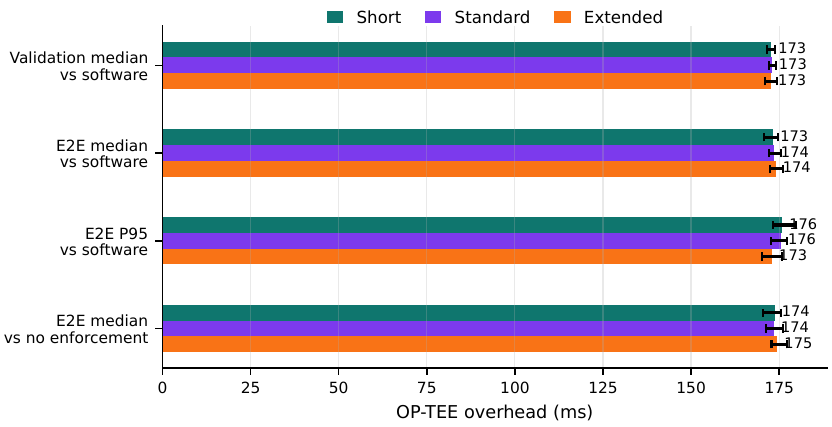}
\caption{Security enforcement overhead of the OP-TEE-backed path. Bars report
means over the seven deadline points for each mission class, and whiskers show
the min--max range.}
\label{fig:security-overhead}
\end{figure}

After establishing trusted-path decision consistency in the degraded suite, we
isolate the cost of moving enforcement into OP-TEE. Fig.~\ref{fig:security-overhead}
shows the OP-TEE-backed overhead relative to software enforcement. Across all
mission--deadline points, OP-TEE increases median validation latency by
172.8\,ms on average, with a min--max range of 171.1--174.5\,ms. It also
increases median end-to-end latency by 173.6\,ms on average and P95 end-to-end
latency by 174.8\,ms on average.

This overhead is stable across mission lengths because it is dominated by the
per-invocation trusted boundary rather than by the number of mission steps. The
result exposes the intended tradeoff: software enforcement provides a
low-latency baseline, while OP-TEE-backed enforcement provides stronger
isolation and auditability at the cost of bounded task-level admission latency
in the emulated environment. In this case, RT-SHCUA therefore keeps high-frequency flight
control outside the secure-world boundary and uses the trusted path for
lower-rate, security-critical skill admission.

\subsection{Evidence Preservation and Audit Cost}

Each allowed, denied, deferred, or fallback-mediated invocation produces a
hash-chain evidence record that links the mission context, runtime state,
enforcement result, and execution outcome. We use a dedicated microbenchmark to
measure the online append cost and offline verification cost of this log.

Table~\ref{tab:evidence-audit-cost} reports the evidence microbenchmark
results. For a 1000-record evidence log, the append path has a median latency of
40.75\,$\mu$s, a P95 latency of 56.46\,$\mu$s, and a P99 latency of
152.88\,$\mu$s. Offline verification costs 13.96\,$\mu$s per record, or about
71.6 thousand verified records per second. These microsecond-level costs are
much smaller than the task-level dispatch and OP-TEE-backed validation
latencies, so evidence preservation does not become the bottleneck of the
admission path. Table~\ref{tab:onboard-footprint} also reports the footprint
of the components that generate and protect the evidence chain.

\begin{table}[!t]
\caption{Evidence append and audit microbenchmark.}
\label{tab:evidence-audit-cost}
\centering
\scriptsize
\setlength{\tabcolsep}{5pt}
\renewcommand{\arraystretch}{1.08}
\begin{tabular}{@{}l r@{}}
\hline
Metric & Value \\
\hline
Evidence-log size & 1000 records \\
Append latency, median & 40.75\,$\mu$s \\
Append latency, P95 & 56.46\,$\mu$s \\
Append latency, P99 & 152.88\,$\mu$s \\
Offline verification cost & 13.96\,$\mu$s/record \\
Verification throughput & 71.6\,k records/s \\
Average record size & 545 bytes \\
\hline
\end{tabular}
\end{table}

\subsection{Summary}

Overall, the evaluation demonstrates bounded task-level mediation for
SHCUA-generated UAV actions. The normal matrix completes 1890 executions without
deadline misses or fallback events, and longer missions keep a stable
per-invocation dispatch cost. The degraded suite achieves 7560/7560 trusted-path
matches against the software reference, confirming correct fallback, defer, and
deny behavior. OP-TEE adds a bounded trusted-admission cost in the emulated
environment, while evidence append and audit remain lightweight. These results
support carrying the current PX4 SITL/Gazebo and QEMU-OP-TEE prototype toward
HITL, hardware TrustZone, and real-flight deployments.

\section{Related Work}

\subsection{LLM-Enabled UAV and Embodied Robotic Agents}

Recent work has increasingly connected large language models (LLMs) and
multimodal models with UAV systems. PX4-based drone-agent studies demonstrate
that natural-language commands and vision-language scene understanding can be
integrated with ROS~2, local models, and real or simulated quadcopters
\cite{lim2025takingflight}. NeLV extends this direction toward multi-scale UAV
mission orchestration, where natural-language instructions are parsed into route
planning, waypoint planning, trajectory execution, and flight monitoring
\cite{yuan2025nelv}. Other UAV-oriented studies explore prompt-driven
multi-drone planning, LLM-to-drone command interfaces over MAVLink/MCP, and
broader LLM-assisted UAV operations and communications
\cite{liu2024promptdrivenmultidrones,ramossilva2026universaldroneinterface,emami2026llmuavsurvey}.
Recent surveys and empirical studies further characterize LLM-assisted UAVs as
an emerging direction for low-altitude mobility, autonomy, mission planning,
swarm coordination, and human--UAV interaction
\cite{tian2025uavsmeetllms,chen2025whenllmsmeetuavs}.
TypeFly further targets the latency side of LLM-driven drone control by
using MiniSpec as a compact intermediate representation, but its goal is
planning efficiency rather than trusted skill admission or degraded
execution~\cite{chen2023typefly}.

These systems show that natural language can serve as a high-level UAV
interface, but they mainly focus on command generation, mission orchestration,
or interface integration. AerialClaw is a highly relevant recent aerial-agent
framework: it organizes UAV skills, PX4/Gazebo or AirSim execution, agent
loops, memory, runtime validation, and edge-cloud planning into an open-source
aerial-agent system \cite{aerialclaw2026framework}. RT-SHCUA is complementary.
Rather than providing a general aerial autonomy framework, it studies the
execution boundary between an SHCUA and a UAV: what information must be attached
to each skill invocation, when the invocation remains admissible, how stale or
unsafe invocations degrade, and how security-critical decisions can be isolated
and audited.

LLM-based robotics has also developed a rich set of techniques for grounding
language in embodied action. SayCan combines LLM task reasoning with affordance
scores from pretrained robot skills so that language proposals are constrained
by what the robot can execute \cite{ahn2022saycan}. LLM-Planner studies
few-shot grounded planning for embodied agents, while Code as Policies and
ProgPrompt use code- or program-like representations to translate
natural-language instructions into executable robot policies or situated task
plans
\cite{song2022llmplanner,liang2022codeaspolicies,singh2022progprompt}.
Inner Monologue uses language feedback from the environment to improve embodied
planning \cite{huang2022innermonologue}, and ChatGPT for Robotics studies
prompt engineering, function libraries, and dialog-based reasoning across
multiple robot platforms, including aerial navigation
\cite{vemprala2023chatgptrobotics}.

More recent embodied and vision-language-action models integrate perception,
language, and control more tightly. PaLM-E incorporates visual and continuous
state inputs into a large language model for embodied reasoning
\cite{driess2023palme}. RT-2 maps visual observations and language instructions
to robot actions as text tokens and transfers web-scale vision-language training
to robotic control \cite{brohan2023rt2}. OpenVLA makes this direction more
accessible through an open-source vision-language-action model trained on large
robot demonstration data \cite{kim2024openvla}. Other work grounds language in
spatial or embodied representations, including EmbodiedGPT, VoxPoser,
Manipulate-Anything, and Voyager
\cite{mu2023embodiedgpt,huang2023voxposer,duan2024manipulateanything,wang2023voyager}.

These works demonstrate that language models can guide embodied tasks through
skills, programs, feedback, and multimodal grounding. However, most of them
focus on task completion, skill composition, manipulation generalization, or
perception-action transfer. UAV control imposes stricter timing, safety, and
authority requirements. RT-SHCUA therefore treats each generated UAV skill not
only as an executable action, but as an admission object carrying deadline
validity, state assumptions, authority scope, fallback behavior, and evidence
requirements.

\subsection{Tool-Use and Computer-Use Agents}

SHCUA-style agents belong to the broader family of LLM-based autonomous agents
that reason, plan, call tools, and interact with external environments. Surveys
of LLM-based agents describe recurring components such as memory, planning,
perception, action, and feedback loops
\cite{wang2023autonomousagents,xi2023riseagents}. ReAct introduced interleaved
reasoning and acting, allowing language models to update their reasoning while
interacting with tools or environments \cite{yao2022react}. Toolformer,
Gorilla, ToolLLM, TaskMatrix.AI, and HuggingGPT extend this direction by
connecting LLMs to APIs, tool repositories, and external models
\cite{schick2023toolformer,patil2023gorilla,qin2023toollm,liang2023taskmatrix,shen2023hugginggpt}.
Recent work also studies long-horizon planning and reliability in large tool
ecosystems and interactive tool-agent-user settings
\cite{yao2024taubench,liu2026planbenchxl}, while surveys summarize tool
learning as a rapidly growing area \cite{qu2024toollearning}.

Computer-use and GUI agents move tool use from API calls to interactive
software environments. Mind2Web, WebArena, VisualWebArena, WebVoyager, and
BrowserGym study web agents operating realistic websites and multimodal web
interfaces
\cite{deng2023mind2web,zhou2023webarena,koh2024visualwebarena,he2024webvoyager,lesellier2024browsergym}.
OSWorld, AndroidWorld, and Windows Agent Arena benchmark agents in real
operating-system or mobile environments
\cite{xie2024osworld,rawles2024androidworld,bonatti2024windowsagentarena}.
WorkArena and WorkArena++ focus on enterprise knowledge-work tasks and
compositional planning \cite{drouin2024workarena,boisvert2024workarenaplusplus},
and Mobile-Agent studies mobile-device control through multimodal perception
\cite{wang2024mobileagent}. Recent OS-agent and computer-using-agent surveys
further systematize this landscape and identify safety and privacy as central
challenges \cite{hu2025osagentsurvey,chen2025cuasafetysurvey}.

These studies are directly relevant to SHCUAs because they show how modern
agents can operate digital interfaces, maintain state, and perform multi-step
workflows. Recent host-acting-agent work further shows that this execution model
has security risks beyond ordinary prompt misuse. Semantic under-specification
can cause an agent to complete a benign goal through risky host-side plans when
process constraints, persistence, exposure, or privilege boundaries are left
implicit \cite{lu2026semanticunderspecification}. TEE-backed SHCUA isolation
then studies how operation classification, authorization, binding, and evidence
generation can be protected for host-level agent actions
\cite{lu2026teebackedisolation}.

RT-SHCUA builds on this computer-use-agent background but changes the execution
domain. Most computer-use agents operate on digital resources, where delayed or
incorrect actions are often recoverable through rollback, user correction, or
task retry. UAVs are different: their state evolves continuously, and an action
can affect a physical vehicle before the SHCUA receives another observation.
RT-SHCUA therefore keeps the flexible tool-use model at the semantic layer but
adds a vehicle-side admission boundary before any agent output can become a UAV
action.

\subsection{Runtime Safety, Security, and Trusted Enforcement}

Safety supervision is a major concern in LLM-based robotics and UAV systems.
SwarmGPT combines high-level LLM-based choreography with a safety filter that
corrects unsafe or infeasible drone-swarm trajectories before deployment
\cite{schuck2024swarmgpt}. SwarmGPT-Primitive further uses motion primitives and
optimization-based filtering to enforce collision, downwash, and actuator
constraints \cite{vyas2024swarmgptprimitive}. These works support the principle
that language-model outputs should be mediated before they determine physical
motion.

A related line of work studies runtime assurance for autonomous and
learning-enabled systems. Runtime assurance systems monitor an advanced or
unverified controller and intervene through a backup controller or safety filter
when necessary \cite{hobbs2021runtimeassurance}. SafeDrones applies runtime
reliability assessment to UAVs through executable dependable identities
\cite{aslansefat2022safedrones}. Other runtime-assurance work demonstrates that
real-time monitors can enforce safety constraints for autonomous spacecraft
inspection \cite{dunlap2023runtimeassurance}, while STPA-based studies analyze
hazards in learning-enabled systems \cite{qi2023stpa}. RT-SHCUA follows the
same separation between high-level autonomy and local runtime mediation, but its
focus is not trajectory correction alone. An SHCUA-generated UAV skill may also
fail because it is stale, state-inconsistent, unauthorized, or outside the
mission contract. The runtime must therefore choose among allow, deny, defer,
and fallback outcomes while preserving both responsiveness and auditability.

LLM agents introduce additional security risks because they combine
natural-language reasoning with tool privileges, memory, and untrusted external
data. Indirect prompt injection shows that untrusted web pages, documents, or
tool outputs can hijack an LLM-integrated application and affect API calls
\cite{greshake2023indirectprompt}. HouYi and InjecAgent study prompt-injection
attacks and benchmarks for tool-integrated agents
\cite{liu2023promptinjection,zhan2024injecagent}. AgentDojo provides a dynamic
benchmark for prompt-injection attacks and defenses, Task Shield checks whether
instructions or tool calls remain aligned with the user task, and ToolEmu
evaluates high-stakes failures of tool-integrated agents
\cite{debenedetti2024agentdojo,jia2024taskshield,ruan2023toolemu}. Recent
surveys and benchmarks, including CUAHarm, emphasize that computer-using agents
create safety risks beyond ordinary chatbot or simple tool-use settings
\cite{deng2024aiagentsunderthreat,tian2025cuaharm,chen2025cuasafetysurvey}.

A second group of recent work focuses on agent memory, control flow, and
architectural isolation. CaMeLs argues for system-level isolation for
computer-use agents, AgentSys uses hierarchical memory management to limit
indirect prompt injection, A-MemGuard defends LLM-agent memory, and recent work
on Zombie Agents and memory-control-flow attacks shows that persistent memory
can become a long-lived steering surface
\cite{foerster2026camels,wen2026agentsys,wei2025amemguard,yang2026zombieagents,xu2026memflow}.
These studies reinforce a key design lesson for RT-SHCUA: the enforcement
boundary should not be reduced to a prompt-level guardrail. It should control
what information crosses the boundary, which action is authorized, and what
execution evidence is preserved.

Closely related to RT-SHCUA are recent studies on OpenClaw-like host-acting agents.
Several works analyze OpenClaw security, including adversarial evaluation,
full-lifecycle defense architecture, runtime defense layers, governance
frameworks, and attack-surface expansion under multi-agent settings
\cite{shan2026clawgrip,ying2026clawguard,li2026openclawprism,ge2026governancearchitecture,wang2026securityopenclaw,jamshidi2026openclawsecurityengineering}.
Semantic under-specification frames the risk that a goal-only instruction can be
translated into a security-divergent plan before any low-level action is
executed \cite{lu2026semanticunderspecification}. TEE-backed SHCUA isolation
moves from semantic risk analysis to protected host-level operation control,
placing security-critical classification, authorization, binding, and evidence
logic in a trusted operation plane \cite{lu2026teebackedisolation}. RT-SHCUA
extends this line of work from host-side computer use to UAV control, where
skill admission must additionally account for vehicle state, task-level
deadlines, degraded recovery, and physical safety.

These risks become more severe when an agent controls a physical platform. UAV
command interfaces can also be sensitive to parameter and command trust: recent
work on ArduPilot shows that legitimate MAVLink messages can still be misused to
manipulate controller behavior when parameter handling and controller-state
validation are insufficient \cite{benchellal2026mavlinkcontrol}. This further
motivates an explicit mediation boundary between agent reasoning and UAV
execution.

Trusted execution environments (TEEs) and isolated execution mechanisms provide
one way to protect security-critical control points. Recent surveys review TEE
secure computation and attestation mechanisms across Intel SGX, Arm TrustZone,
AMD SEV, and RISC-V TEEs \cite{li2023teesurvey,menetrey2022attestation}.
Fortress uses TEEs to protect IoT peripheral data paths
\cite{yuhala2023fortress}, and TrustZone-based PLC work evaluates
OP-TEE-style isolation for industrial control logic \cite{li2024plctrustzone}.
Recent TrustZone/OP-TEE work also studies runtime integrity measurement and
attestation inside the TEE \cite{mao2025pdrima}. RT-SHCUA applies this
trusted-enforcement idea to UAV skill admission: authorization, freshness,
command binding, fallback authorization, and evidence digests can be protected
without moving the full SHCUA or high-frequency flight-control loop into the
trusted boundary.

\subsection{Comparison with Existing Work}

Table~\ref{tab:related_work_comparison} summarizes the relationship between
representative prior work and RT-SHCUA. Existing UAV-LLM systems show that
language models can generate commands, mission plans, and aerial-agent
workflows. Embodied-agent and VLA work shows that language can be grounded in
robot skills and perception. Tool-use and computer-use agents demonstrate
general interfaces for software automation. Recent SHCUA and OpenClaw security
studies analyze semantic under-specification, prompt-injection risks, runtime
defense, memory security, and TEE-backed confinement on the host side. Runtime
assurance, safety filtering, agent-security, and TEE work provide important
mechanisms for local intervention, action control, and protected enforcement.

The remaining gap is that these directions are usually studied separately.
RT-SHCUA treats SHCUA-to-UAV control as a single mediated execution problem. It
restructures free-form agent tool use into deadline-aware, state-aware,
security-enforced, fallback-capable, and evidence-preserving UAV skill
admission.

\begin{table*}[!t]
\caption{Comparison with representative existing work.}
\label{tab:related_work_comparison}
\centering
\scriptsize
\setlength{\tabcolsep}{1.8pt}
\renewcommand{\arraystretch}{1.16}
\begin{tabularx}{\textwidth}{@{}
>{\raggedright\arraybackslash}p{2.05cm}
>{\raggedright\arraybackslash}p{3.10cm}
>{\centering\arraybackslash}X
>{\centering\arraybackslash}X
>{\centering\arraybackslash}X
>{\centering\arraybackslash}X
>{\centering\arraybackslash}X
>{\centering\arraybackslash}X
@{}}
\hline
\textbf{Work category} &
\textbf{Representative works} &
\textbf{Physical interface} &
\textbf{Deadline admission} &
\textbf{Fallback path} &
\textbf{Security / accountability} &
\textbf{Trusted isolation} &
\textbf{SHCUA restructuring} \\
\hline

Natural-language UAV control
& \cite{lim2025takingflight,yuan2025nelv,tian2025uavsmeetllms,chen2025whenllmsmeetuavs,emami2026llmuavsurvey,ramossilva2026universaldroneinterface}
& Yes & Partial & Partial & Partial & No & No \\

Agentic aerial frameworks
& \cite{aerialclaw2026framework}
& Yes & Partial & Partial & Partial & No & Partial \\

Embodied robotic agents and VLAs
& \cite{ahn2022saycan,liang2022codeaspolicies,singh2022progprompt,driess2023palme,brohan2023rt2,kim2024openvla}
& Yes & Partial & Partial & Partial & No & No \\

Tool-use and computer-use agents
& \cite{yao2022react,schick2023toolformer,liang2023taskmatrix,patil2023gorilla,qin2023toollm,xie2024osworld,rawles2024androidworld}
& Partial & No & Partial & Partial & No & No \\

Computer-use agent safety
& \cite{foerster2026camels,tian2025cuaharm,chen2025cuasafetysurvey,hu2025osagentsurvey}
& Partial & No & Partial & Yes & Partial & Partial \\

Host-side SHCUA / OpenClaw security
& \cite{lu2026semanticunderspecification,lu2026teebackedisolation,shan2026clawgrip,ying2026clawguard,li2026openclawprism,wang2026securityopenclaw}
& Partial & Partial & Partial & Yes & Partial & Partial \\

Low-latency planning and compact representations
& \cite{chen2023typefly,liu2026planbenchxl}
& Partial & Yes & Partial & Partial & No & No \\

Safety supervision and runtime assurance
& \cite{schuck2024swarmgpt,vyas2024swarmgptprimitive,hobbs2021runtimeassurance,aslansefat2022safedrones,dunlap2023runtimeassurance}
& Yes & Partial & Yes & Partial & No & No \\

Agent security and prompt-injection defense
& \cite{greshake2023indirectprompt,liu2023promptinjection,zhan2024injecagent,debenedetti2024agentdojo,jia2024taskshield,ruan2023toolemu}
& Partial & No & Partial & Yes & Partial & No \\

TEE and isolated enforcement
& \cite{li2023teesurvey,menetrey2022attestation,yuhala2023fortress,li2024plctrustzone,mao2025pdrima}
& Partial & Partial & Partial & Yes & Yes & No \\

\textbf{This work}
& --
& \textbf{Yes} & \textbf{Yes} & \textbf{Yes} & \textbf{Yes} & \textbf{Yes} & \textbf{Yes} \\
\hline

\multicolumn{8}{@{}p{\textwidth}@{}}{\scriptsize
``Yes'', ``Partial'', and ``No'' indicate whether the property is a first-class
design objective of the corresponding line of work, rather than whether an
individual implementation contains an isolated mechanism.}
\end{tabularx}
\end{table*}

\section{Conclusion}

This paper studies how an SHCUA can be adapted to natural-language UAV control without placing slow and open-ended language reasoning inside the real-time flight-control loop.
The core solution is to restructure SHCUA influence over UAV behavior.
Instead of directly executing agent-generated commands, the system compiles task-level agent decisions into contract-bound UAV skill invocations.
Each invocation carries temporal validity, state constraints, authorization scope, fallback semantics, and evidence requirements.

The proposed architecture separates semantic reasoning, mission compilation, onboard real-time execution, and security/safety enforcement.
It also supports degraded execution and trusted or isolated enforcement for critical control points.
In this way, the system preserves the usability of natural-language UAV control while bounding the authority of the SHCUA and maintaining real-time responsiveness, security enforcement, auditability, verifiability, and traceability.

\section*{Acknowledgments}
The authors would like to thank the editor-in-chief, associate editor, and reviewers for their valuable comments and suggestions. This research was supported by the National Natural Science Foundation of China (62232013, U24A20243, 62572377, 62302363), the Innovation Capability Support Program of Shaanxi (No. 2023-CX-TD-02), the Xidian University Specially Funded Project for Interdisciplinary Exploration (No. TZJHF202502) and the Fundamental Research Funds for the Central Universities (No. ZDRC2202).


\bibliographystyle{IEEEtran}
\bibliography{refs}

\end{document}